\documentclass[a4paper,10pt]{article}
\usepackage[dvips]{graphicx}

 \usepackage{amsmath,amsfonts,amssymb}
\usepackage{color}
 \usepackage{listings}

\definecolor{colKeys}{rgb}{0,0,1}
\definecolor{colIdentifier}{rgb}{0,0,0}
\definecolor{colComments}{rgb}{0,0.5,1}
\definecolor{colString}{rgb}{0.6,0.1,0.1}
\title{Cameleon language\\Part 1~: Processor}
\author{O. Cugnon de Sevricourt and V. Tariel\footnote{List of authors by alphabetical order}}

\begin{document}

\maketitle
\begin{abstract}
Emergence is the way complex systems arise out of a multiplicity of relatively simple interactions between primitives. An example of emergence is the Lego game where the primitives are plastic bricks and their interaction is the interlocking.  Since programming problems become more and more complexes and transverses,our vision is that application development should be process at two scales~: micro- and macro-programming where at the micro-level the paradigm is step-by-step and at macro-level the paradigm is emergence. For micro-programming, which focuses on \textit{how things happen},  popular languages, Java, C++, C, Python, [1], are imperative writing languages where the code is a sequence of sentences executed by the computer. For macro-programming, which focuses on \textit{how things connect}, popular languages, labVIEW, Blender, Simulink, Quartz Composer, Houdini are graphical data flow languages such that the program is a composition that is a directed graph with operators, a unit-process consuming input data and producing output data, and connectors, a data-flow between an output data and an input data of two operators. However, despite their fruitful applications, these macro-languages are not transversal since different data-structures of native data-structures cannot be integrated in their framework easily. Cameleon language is a graphical data flow language following a two-scale paradigm. It allows an easy \textit{up-scale} that is the integration of any library writing in C++ in the data flow language. This integration requires a simple registration of \textit{data}, data dictionary, \textit{input}, the ways to get data, \textit{output}, the ways to display/save data, \textit{operator}, operator dictionary. Cameleon language aims to democratize macro-programming by an intuitive interaction between the human and the computer where building an application based on a data-process and a Graphical User Interface (GUI) is a simple task to learn and to do. Moreover, Cameleon language allows conditional execution and repetition to solve complex macro-problems. In this paper we introduce a new model based on the extension of the petri net model for the description of how the Cameleon language executes a composition. In two companion papers we will present the Cameleon modern architecture and the up-scale feature. 
%it is challenge to have an macro(over)-view and micro(deep)-view on it.
% because the condition of execution and the actualization of the state of the input/output data after an execution can be tuned.   in including the data-transfer
% In order to solve efficiently these problems, an idea is to operate the inverse process of emergence by scale separation: decomposition in term of simple problems with their interactions. A challenge is the definition of an interpreted language for the representation of this inverse problem. A data-flow programming language is a graphical programming language allowing the creation of a composition like a Lego game given a directed graph of the data flowing between operations. Many languages, as Labview, blender, exist but they are restricted to a specific application because they do no include the concept of data structure abstraction.
%  It is our goal to create a language that will offer developers to translate easily their library, a collection of operations applied on data-structures, the corresponding data-flow programming. In this paper, we will explain the framework of our data-flow language using a modified petri net. In a companion paper, we will present the utilisation of the paradigms of a modern language with a appropriate architecture in order to reach the abstract data flow programming. 
\end{abstract}

\section{Introduction}
Programming language is the system of communication between a human and a computer having a Central Processing Unit (CPU). A CPU is
\begin{itemize}
\item  a calculator executing the loops, the logical and arithmetic tests,
\item  a controller commanding the devices and the data-transfer between their where the devices can be computer data storages, screen. 
\end{itemize}
The behavior of the CPU is defined by its state, $s_{t}$ at time $t$, and its evolution, $s_{t+1}$, is governed by a Markovian deterministic function, $f$, depending on its previous state, $s_t$, and the instruction $it_t$ at time $t$~:
\[ s_{t + 1} = f(s_t,i_t).\] 
At each clock tick, this function is applied. The CPU works at a given clock rate given the number of clock per second.%, for instance, for a Intel Pentium 4, this clock rate is 3 GHz (three billion clocks/second). 
The set of instructions is:
\begin{itemize}
 \item  arithmetic instructions such as addition, subtraction,
 \item  logic instructions such as and, or, not,
 \item  data instructions such as move, load,
 \item  control flow instructions such as goto, if.
\end{itemize}
These instructions are encoded in machine language, i.e. a sequences of bits: $\{0,1\}^n$ where n is the number of bits of the computer-architecture. In this language, a code is a list of instructions that will be execute by the CPU without transformation. 
At this level, programming  requires a look-up table to have the instruction in machine language and a deep knowledge of the CPU architecture. Obviously, this human/computer communication is not efficient. Therefore programmers create some languages to encode a set of instruction efficiently and implement some compilers to convert these languages in the machine language. It emerged the assembly languages where the code, a set of instructions, can be read and written by a programmer. These simple instructions are then assembled directly into machine code. First, the language is specific to a particular CPU architecture family and, second, the writing of simple programs, as counting the number of a specific character in a given text file, requires a long list of instructions. The abstraction between the language and the machine language leads to the middle- and high-level language. On the one hand, these languages can be textual with a syntax composed by a lexical analysis, the conversion of a sequence of characters into a sequence of tokens, and a syntactic analysis, the determination of the grammatical structure of the sequence of tokens. Nowadays, the grammar of the popular languages, Java, C++, C, [1], is imperative such that the program is is a sequence of sentences where each sentence is a command to be performed by the computer. On the other hand, these languages can be  graphical. In this category, the most popular languages are LabVIEW, Blender, Simulink, Quartz Composer, Houdini, Grafcet. Most of them are data flow programming languages, such that the program is a composition of data-process and GUI. The data process is a directed graph with operators, a unit-process consuming input data and producing output data, and connectors, a data-flow between an output data and an input data of two operators. An operator can be either a composition or a compiled  function (procedure) writing in textual language.  GUI includes data input/output and operator controllers. These language are closed to their specific domain of applications since  different  data structures of the native data structures cannot be integrated easily.\\
Cameleon language is a new open-source data flow programming languages in a multi-scale programming paradigm. At a micro-scale, programmers code algorithms in C++ in a analytical way. They must know everything about what happen. For this tactical job, it appears that the most suitable solution is an imperative writing language.  At a macro-scale, programmers  code in a phenomenological way. The data-process is the combinaison of unit-processes(operators) where each operator can be seen as a black box consuming/producing resources. For this strategical job, programmers use the Cameleon language to create an application with drag-and-drop for the creation of the data-process and the graphical user interface containing the data input/output and the operator controllers. Two-scale paradigm enforces a programmer  to think about data-process in a divide and conquer paradigm~: micro-scale programming implements the elementary operators for the data-process at macro-scale programming. In the Cameleon language, we can focus on both scale since the up-scale is an easy and fast job. It is just the registration of a dictionary containing \textit{data}, data dictionary , \textit{input}, the ways to get data, \textit{output}, the ways to display/save data, \textit{operator}, operator dictionary. This development pipeline makes programming efficient at both scale. Moreover at macro-scale, everybody can be a computer scientist for the development of end-user applications.\\
Petri net is a powerful model for a mathematic discretion of a language. Like industry standards such as UML activity diagrams, BPMN and EPCs, it offers a graphical notation for stepwise processes that include choice, iteration, and concurrent execution. Unlike these standards, Petri nets have an exact mathematical definition of their execution semantics, with a well-developed mathematical theory for process analysis. However,  simple repetition as loop or conditional execution as if/else requires the composition of a large number of elements. Obviously, an interpreted language based on it is not efficient. To overcome that, we introduce a new extended petri net model where we can tune the execution rules in order to design repetition and conditional execution in a simple way.   
The formal semantics of the language given by this extended petri net model is directly interpreted by the execution loop. In this article, we will define the execution semantic of the Cameleon language.
% In programming language theory, semantics is the field concerned with the rigorous mathematical study of the meaning of programming languages and models of computation. The formal semantics of a language is given by a mathematical model that describes the possible computations described by the language. 
% 
% Direct translation between this model and the interpreted language 
% 
% 
% apllication is the integration of  to include some input/output to  
% 
% problem to mix the scale to force the multi-scale paradigm.
% high-level imperative language  programming language as Python, C++ with Qt library. 
% the data-process 
%  to manage. But, it seems easier to make this combinaison with a data-flow programming language.  
% 
% Application software

%  is contrasted with system software and middleware, which manage and integrate a computer's capabilities, but typically do not directly apply them in the performance of tasks that benefit the user. The system software serves the application, which in turn serves the user.
% 
%   this kind of data in ignoring where the more important is the connection between the unit-processes. This multi-scale programming paradigm can be applied when the computer program is complex. Moreover, this language democratize the programmation since the creation of a composition is easy as a lego game, just the drag-and-drop  of unit-processes and connect their.  

% Open-source,
% \section{Cameleon engine}
\section{Extended petri net}
% These condition rules are those implemented in the cameleon language. 
%  This model is turing complete. 
% In this part, I will follow the orgamization of the chapter 1 of the Peterson's book <ref>Peterson, James Lyle (1981), Petri Net Theory and the Modeling of Systems, Prentice Hall</ref>
% 
% why we do that??? answer this question
\subsection{Composition}
\subsubsection{Structure}
The petri structure is  composed of four parts: a set of ''data``, $D=\{d_i\}_{0\leq i<n}$, a set of ''operators``, $Op=\{op_i\}_{0\leq i<k}$', an ''input function'' $I$, and an ''output function'' $O$. The input function represents the inner data flow of each operator that is a mapping from an operator, $op$ to a collection of data,  $I(op)=\{d_0,\ldots,d_n\}$. The same for the output function except input becomes output. For instance, we can have~:
\begin{equation}
C_0 = (D_0,Op_0,I_0,O_0)\text{ with }\left\{\begin{array}{*{1}{l}} 
D_0=\{d_0,d_1,d_2,d_3,d_4,d_5,d_6\} \\
O_0=\{op_0,op_1,op_2,op_3\}\\
I_0(op_0)=\{d_0,d_1\}\text{, } I_0(op_1)=\{d_2\}\text{, } I_0(op_3)=\{d_2\}\text{, } I_0(op_3)=\{d_4,d_5\},\\
O_0(op_0)=\{d_2,d_3\}\text{, } O_0(op_1)=\{d_4\}\text{, } O_0(op_2)=\{d_5\}\text{, } O_0(op_3)=\{d_6\}
\end{array}\right. 
\end{equation}
\begin{equation}
C_1= (D_1,Op_1,I_1,O_1)\text{ with }
\left\{\begin{array}{*{1}{l}} 
D_1=\{d_0,d_1,d_2,d_3,d_4,d_5,d_6,d_7,d_8,d_9\} \\
O_1=\{op_0,op_1,op_2,op_3,op_4,op_5\}\\
I_1(op_0)=\{\}\text{,  } I_1(op_1)=\{d_0,d_1\}\text{, } I_1(op_2)=\{d_3,d_8\},\\
I_1(op_3)=\{d_2,d_4\}\text{,  } I_1(op_4)=\{d_5,d_6\}\text{, } I_1(op_5)=\{d_7\},\\
O_1(op_0)=\{d_1\}\text{,  } O_1(op_1)=\{d_2\}\text{,   } O_1(op_2)=\{d_4\},\\
O_1(op_3)=\{d_5,d_6\}\text{,  } O_1(op_4)=\{d_7,d_9\}\text{,   } O_1(op_5)=\{d_8\}
\end{array}\right.
\end{equation}
In our model, a data  cannot be included in the inner data flow and in the outer data-flow of the same operator~:
\[
 \forall op\in Op:\quad I(op)\cap O(op)=\emptyset .
\]
We add this consdition to avoid a conflict with the marking function update, see subsection~\ref{subsub:update}.
% This condition ensures there is no conflict during the update 
% \begin{itemize}
%  \item the number of input arc and the number of output arcs for a data  is inferior or equal to 1, this condition allows a simple condition to update the marking 
% \item   
% \end{itemize}
\subsubsection{Graph}
\begin{figure}
\begin{center}
\begin{tabular}{cc}
$C_0$=&\includegraphics[height=4cm]{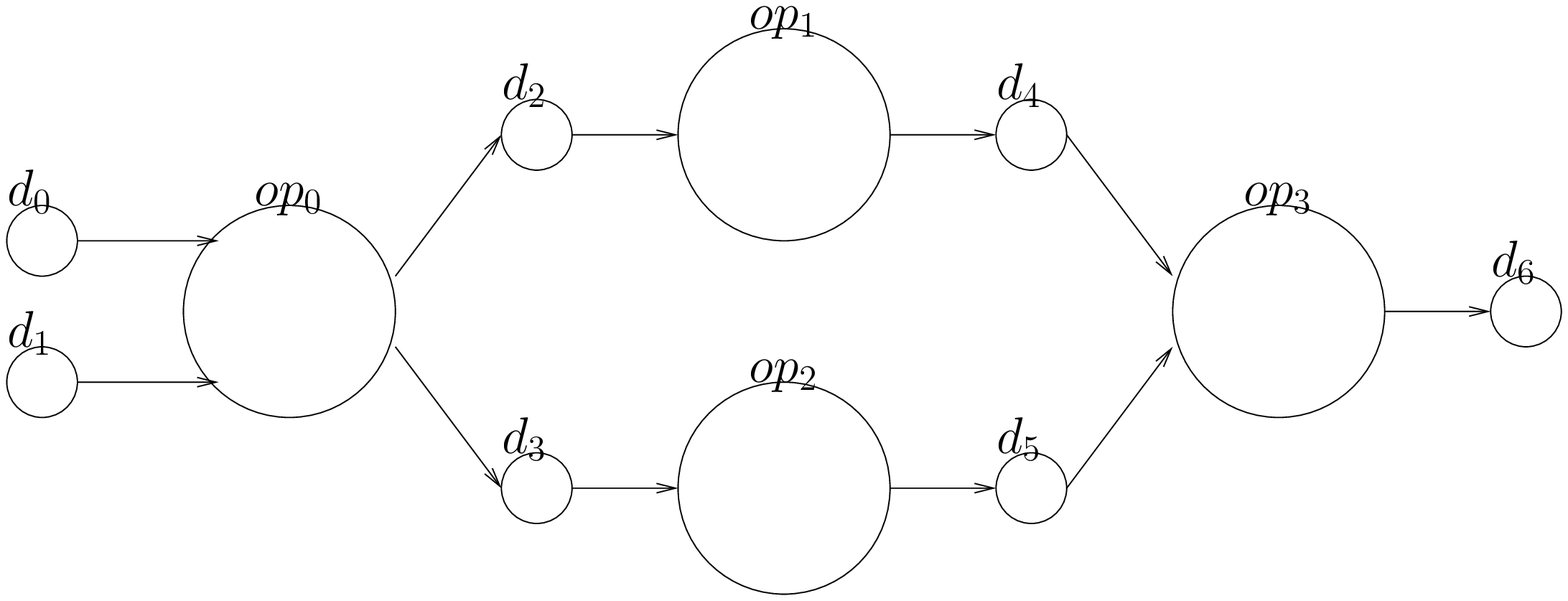}\\
 $C_1$=&\includegraphics[height=4cm]{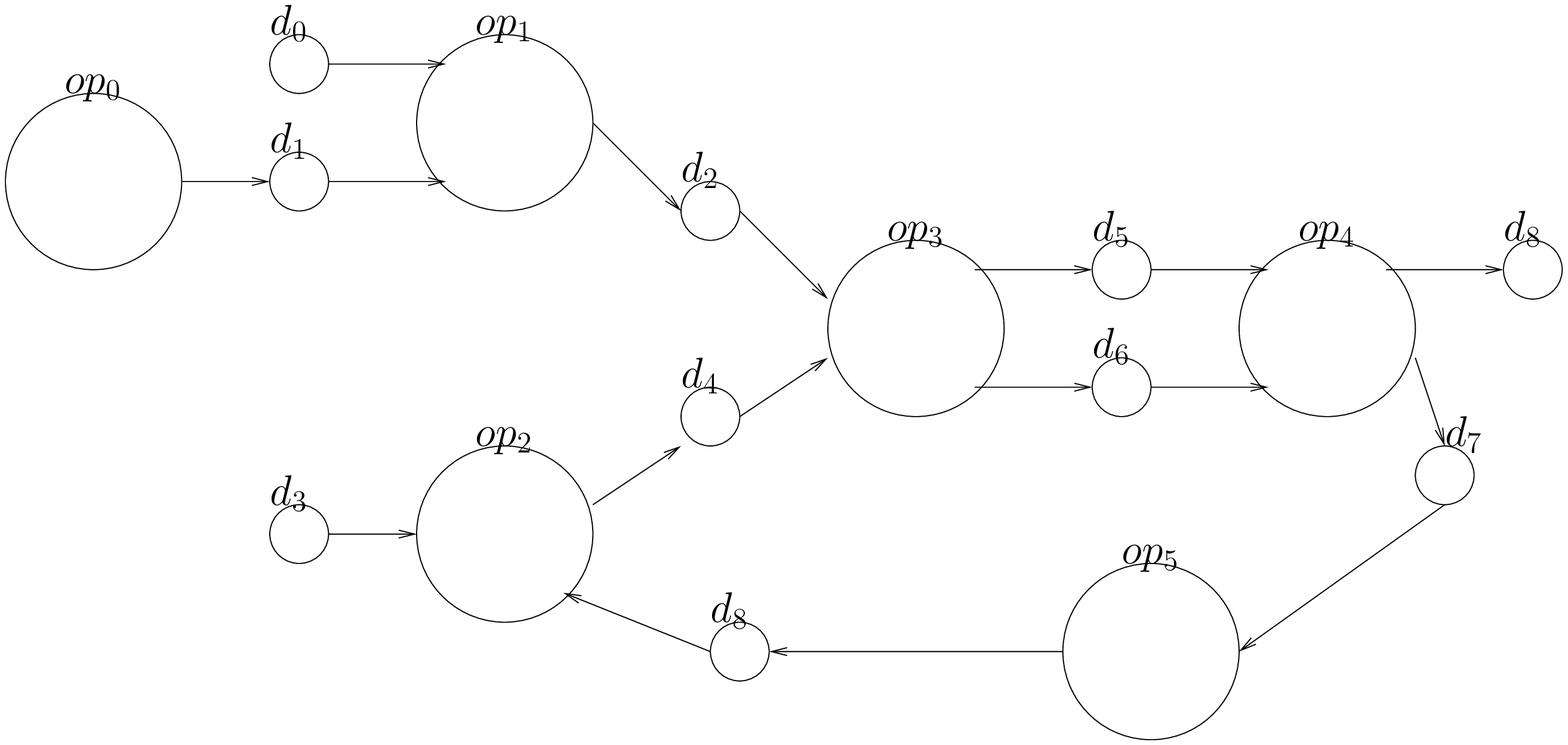}
\end{tabular}
\caption{Graphical representation of the Petri net}
\label{fig:graphpetri}
\end{center}
\end{figure}

The graphical representation of this structure $(D,Op,I,O)$ is a bipartite directed graph. This bipartite directed graph is an ordered pair $D = (V, A)$ with~:
\begin{itemize}
 \item $V$ a set whose elements are called vertices divided in two disjoint sets $D=(d_i)_{0\leq i <n}$ and $Op=(op_i)_{0\leq i <m}$,
\item $A$ a set of ordered pair, whose elements are called arcs, connecting a vertex of $D$ to one of $Op$ or the opposite, where $A = \{(d,op):d\in I(op)\}\cup \{(op,d):d\in O(op)\}$ 
\end{itemize}
A small circle represents a data element and a big circle an operator element. The figure~\ref{fig:graphpetri} shows the graphical representation of the previous petri net structures.
\subsubsection{Marking}
\begin{figure}
\begin{center}
\includegraphics[height=4cm]{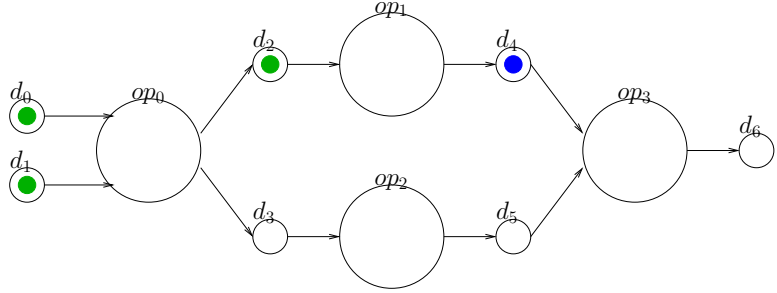}
\caption{Graphical representation of the Petri net with its marking function}
\label{fig:graphmarking}
\end{center}
\end{figure}

The petri net structure is static. The dynamic of this model is given by the successive execution operators that set the information contained in the data.  In our extension, this information can have three states~: void, new and old. The marking function $\mu^t$ returns this state at time $t$ where $\mu^t(d)=2$ for a new token in the data $d$, $\mu^t(d)=1$ for an old token, $\mu^t(d)=0$ for a void token. The petri net is a structure and a marking function $(C,\mu^t)$. For the graphical representation, a blue point represents a new token,  a green point an old token and no point a void token. For instance, this marking function~:
\[
\mu^t(d_0)=1,\mu^t(d_1)=1,\mu^t(d_2)=1,\mu^t(d_2)=1,\mu^t(d_3)=0,\mu^t(d_4)=2,\mu^t(d_5)=0,\mu^t(d_6)=0 
\]
and the petri net $C_0$ are represented by the figure~\ref{fig:graphmarking}.
\subsection{Execution rules}
The execution rules define~:
\begin{enumerate}
\item the condition to allow the execution of an operator,
\item the process  on the input data information  to produce the  output data information done by the operator execution,
\item the update of the marking function of the input/output data of an operator after its execution.   
\end{enumerate}
We can tune each rule to allow~:
\begin{enumerate}
\item the creation of a dictionary of operators for each data-process field for the rule (2),
\item the design of compositions containing conditional executions and repetition in a simple way for the rules (1), (2) and (3). In the section~\ref{sec:compose}, we will design pattern compositions for if/else and loop.
\end{enumerate}
%   The tune of the operator process allows .  The tune of the update and the condition . 
% The first part presents how the process modifies the information contained in the datas.  the general case for the most used execution rules in practice and the second part introduce the mathematical tools to take into account of the specific cases. 
\subsubsection{Execution condition}
\begin{figure}
\begin{center}
\includegraphics[height=1.5cm]{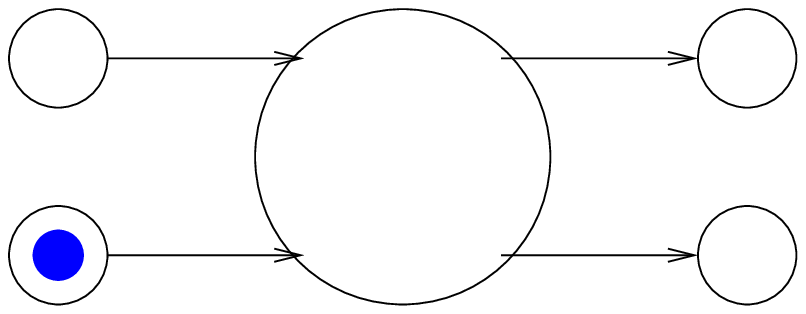}
\includegraphics[height=1.5cm]{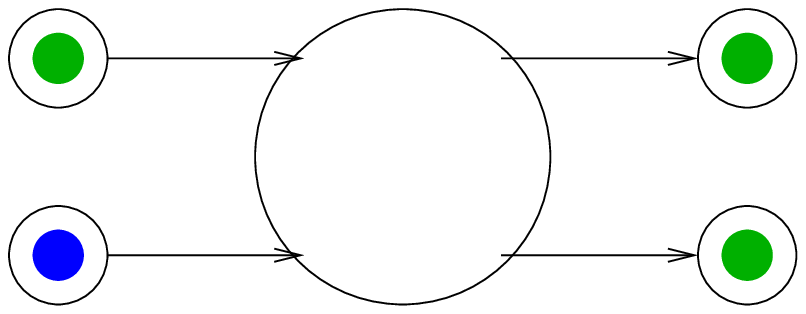}
\includegraphics[height=1.5cm]{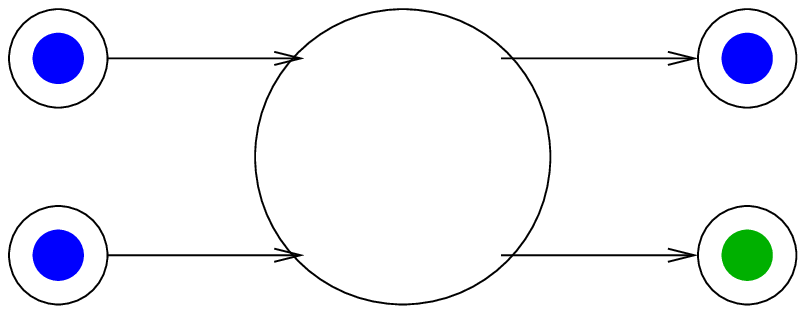}
\caption{Left operator cannot be executed because one input data does not have a token, center operator can be executed because each input data has a token and at least one with new and each output data  does not contain a new token, right operator cannot be executed because one output data  contains a new token }
\label{fig:executeopetaor}
\end{center}
\end{figure}

Depending on token states in the input/output data, a operator can(not) be execute. A predicate execution function, $e_{op}$, returns true if the operator can be executed and false otherwise. In the general case, an operator, $op$, can be executed at time $t$ when
\begin{enumerate}
 \item each input data  contains a new or old token,
\item at least one input data  contains a new token,
\item each output data  does not contain a new token.
\end{enumerate}
The general predicate execution function is~:
\[
 e_{op}(I,0,\mu^t) = \left\{
  \begin{array}{l l}
    1 & \text{for}\quad \forall d\in I(op):\mu^t(d)>0\quad \text{and}\quad\exists d\in I(op):\mu^t(d)=2 \quad \text{and}\quad\forall d\in O(op):\mu^t(d)< 2\\
    0&  \text{otherwise}\\
  \end{array} \right.
\]
The figure~\ref{fig:executeopetaor} shows some configurations where a operator can(not) be executed.
% For if/else composition pattern and the loop composition pattern, we introduce two special predicate execution functions. The first one, $e_{s}$, called synchone, returns true for each input data  contains a new  token and each output data  does not contain a new token, and false otherwise~:
% \[ e_{s}(op,I,0,\mu^t) = \left\{
%   \begin{array}{l l}
%     1 & \quad \text{for}\quad \forall d\in I(op):\mu^t(d)=2\quad \text{and}\quad \forall d\in O(op):\mu^t(d)< 2\\
%     0& \quad \text{otherwise}\\
%   \end{array} \right.
% \]
% The second one, $e_{m}$, called merge, returns true for at least one input data contains a new  token and each output data  does not contain a new token and false otherwise~:
% \[
%  e_{m}(op,I,0,\mu^t) = \left\{
%   \begin{array}{l l}
%     1 & \text{for}\quad \exists d\in I(op):\mu^t(d)=2 \quad \text{and}\quad\forall d\in O(op):\mu^t(d)< 2\\
%     0&  \text{otherwise}\\
%   \end{array} \right.
% \] 
\subsubsection{Process}
An access function, noted $a^t$, returns the information contained in data at time $t$.
The operator process sets the information contained in its output data, $(a^{t+1}(d^o_0),\ldots,a^{t+1}(d^o_n))$ depending on the information contained in its input data, $(a^{t}(d^i_0),\ldots,a^{t}(d^i_p))$. The process function $f_{op}$ defines that as follows~:
\[
\left(a^{t+1}(d^o_0),\ldots,a^{t+1}(d^o_n)\right) = f_{op}\left(a^{t}(d^i_0),\ldots,a^{t}(d^i_p) \right)
\]
where $op$ is the operator executed at time $t$.
For instance, the operator of type $<$ has two input number data, $d^i_{0}$, $d^i_{0}$ and one output boolean data $d^o$. At time $t$, its execution produces an information in  its output data as follows~:
\[
a^{t+1}(d^o) =  \left\{ \begin{array}{l l}
    \text{true} & \text{for}\quad a^t(d^i_{0})<a^t(d^i_{1})\\
    \text{false}&  \text{otherwise}
  \end{array} \right.
\] 
\subsubsection{Update}
\label{subsub:update}
\begin{figure}
\begin{center}
\includegraphics[height=1.5cm]{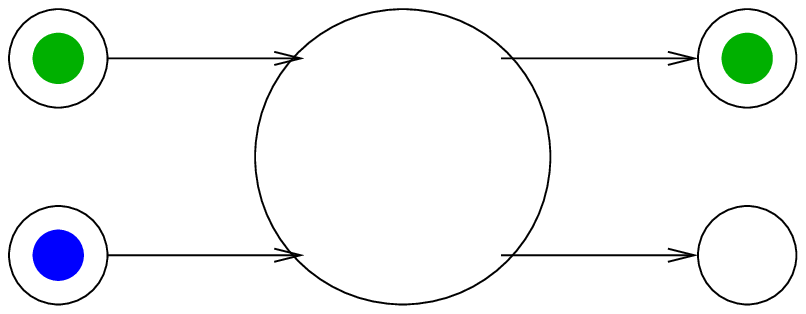}
\includegraphics[height=1.5cm]{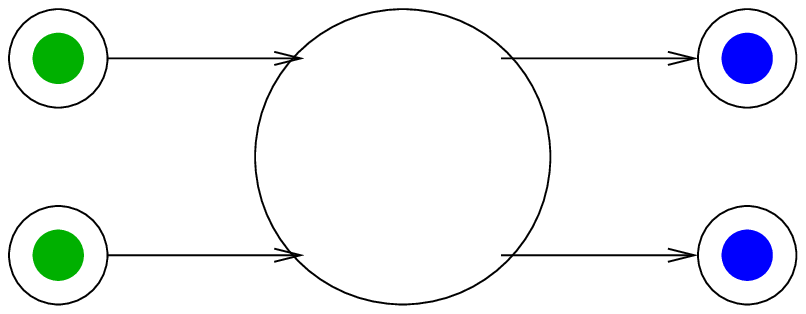}
\caption{Left operator is before the execution, right operator after the execution}
\label{fig:updateopetaor}
\end{center}
\end{figure}

After the execution of an operator executed at time $t$, an update function, $u_{op}$, sets the token states of these input/output data as follows $\mu^{t+1}= u_op(\mu^{t})$.  In the general case, the conditions are~:  
\begin{enumerate}
 \item the tokens in the input data  are set to old, 
\item  the tokens in the output data  are set to new.
\end{enumerate}
that is~:
\[\text{for } \mu^{t+1}(d)= \left\{
  \begin{array}{l l}
     2 & \quad \text{for}\quad d \in O(op)\\
     1& \quad \text{for}\quad d \in I(op) \\
    \mu^t(d) & \quad \text{otherwise}\\
  \end{array} \right.
\]
This function is ill-defined for a data belonging to the input data and the output data of the same operator. However, we cannot have this case since we add a condition in the Petri net structure. The figure~\ref{fig:updateopetaor} shows one example. 

% For if/else composition pattern and the loop composition pattern, we introduce two special update function. The first one, $u_{m}$, called merge, put on old token in the input data containing a new token previously and put a new token in its output data. The second one, $u_{if/else}$, called if/else,  
% 
% 
% 
% 
% 
% For instance, a operator with an if-else type, has one input data, $d$, contening a boolean data, $d_i=0\text{ or }1$,  and two output datas, $d_{o,if}$ and  $d_{o,else}$, has an update post-execution function defined as follows~:
% \[\text{for }\beta(op)=if/else,\quad   \mu^{t+1}(d)= \left\{
%   \begin{array}{l l}
%      2 & \quad \text{for}\quad d = d_{o,if} \text{ and } a^t(d_i)=1 \\
%      2 & \quad \text{for}\quad d = d_{o,else} \text{ and } a^t(d_i)=0 \\
%     1 & \quad \text{for}\quad d = d_{i} \\
%     \mu^t(d) & \quad \text{otherwise}\\
%   \end{array} \right.
% \]
\subsection{Processor}
% \subsubsection{Execution unit}
\subsubsection{Execution by execution}
At each step of time, the processor executes an operator until convergence. An operator execution is the application of its process function  and its update function. At the initial state, the marking function is set by the user manually. The operator to be executed at time $t$ is selected among all operators that can be executed~:
$\{\forall op \in Op:  e_{op}(I,0,\mu^t)=1 \}$. For a void set, we reach the convergence. Otherwise, we select the operators, $op_e$ that wait to be executed for the longest time~:$
 \arg_{\forall op \in Op}\min    \{t':   \forall t''\in [t',t]\quad   e_{op}(I,0,\mu^{t''})=1 \}$.
Because more than one operator can minimize this time, for a deterministic selection, we pick-up the operator having the minimum index in making the assumption that the operators are ordered. In the next section, we will apply this processor for if/else and loop composition patterns. 
\subsubsection{Concurrent execution}
In concurrent execution, many operators are executed simultaneously. 
% As said previously, the operator execution is decomposed in two steps~: process function and update function. The process function consume and produce the information in the data.
To avoid a critical section in the access of the information in the data with the process function and in the access of the tokens in the update function, we do not allow the execution of two neighborhoods operators.  In a continuous time $t$, we introduce the set $E^t=\{op_{i_1},\ldots, op_{i_k}\}$ representing all operators running. When an operator finishes executing,  it is removed from this set and this procedure is applied to start the execution that can be run~:
\begin{enumerate}
 \item while $A=\{op  : e_{op}(I,0,\mu^t)=1\text{ and } \left(I(op)\cup 0(op)\right)\bigcap \left(\cup_{\forall op'\in E^t} (I(op')\cup O(op') ) \right) =\emptyset \}\neq \emptyset $
\begin{enumerate}
 \item extract $op$ from $A$ waiting the longest time,
 \item push it in the list $E^t$ and start its execution. 
\end{enumerate}  
\end{enumerate}
% \[
%  op = \arg\min \{op  : e_{op}(I,0,\mu^t)=1\text{ and } (I(op)\cup 0(op))\bigcap \left(\cup_{\forall op'\in E^t} (I(op')\cup O(op') ) \right) =\emptyset \}
% \]
% continious time with event
% In concurrent programming a critical section is a piece of code that accesses a shared resource (data structure or device) that must not be concurrently accessed by more than one thread of execution
% continious time
% This brute-force approach can be improved upon by using semaphores. To enter a critical section, a thread must obtain a semaphore, which it releases on leaving the section. Other threads are prevented from entering the critical section at the same time as the original thread, but are free to gain control of the CPU and execute other code, including other critical sections that are protected by different semaphores
%  If the processor execute the selected operator,th
% \paragraph{Input/Output tokens}
% From outside, an  user can add/remove tokens in the composition.  In this paper, this function is only used for the initialisation of the composition. However, in practice, an user interacts throughout the process.
% \subsubsection{Evolution}
\section{Composition pattern}
\label{sec:compose}
In our extended Petri net, the execution of a composition depends on two scales. At macro-scale we have the connection between the operators and the data and at micro-scale we have the execution rules of each operator.  In this section, we will begin with the execution rules of the operators required by the compositions.  
\subsection{Operators}  
\subsubsection{If/else}
The operator if/else has~:
\begin{itemize}
 \item two input data, $d^i_0$ and $d^i_1$ where the second input data contains a boolean information and two output data $d^o_{if}$ and $d^o_{else}$,  
\item a predicate execution function as the general case,
\item a process function that copies the first input data to the first output data $a^{t+1}(d^o_{if})=a^t(d^i_0)$ for $a^{t}(d^i_{1})$ equal to true and  the second one otherwise,
\item an update function that sets at old  all tokens in the input data and set at new the first token in the output data $d^i_{if}$ for  $a^t(d^i_1)$ equals to true and the second one otherwise.
\end{itemize}
\subsubsection{Merge}
The operator merge has~:
\begin{itemize}
 \item two input data, $d^i_0$ and $d^i_1$ and one output data $d^o$,  
\item a predicate execution function that returns true for at least one input data containing a new token and each output data  does not contain a new token and false otherwise,
\item a process function that copies the input data containing the new token to the output data,
\item an update function that sets at old the token in the input data containing the new token previously and at new the token in the output data.
\end{itemize}
\subsubsection{Synchrone}
The operator synchrone has~:
\begin{itemize}
 \item two input data, $d^i_0$ and $d^i_1$ and two output data, $d^o_0$ and $d^o_1$  
\item a predicate execution function that returns true for all input data containing a new token and each output data  does not contain a new token and false otherwise,
\item a process function that copies the input data to the output data, $a^{t+1}(d^o_{0})=a^t(d^i_0)$ and $a^{t+1}(d^o_{1})=a^t(d^i_1)$,
\item an update function as the general case.
\end{itemize}
\subsubsection{Increment}
The operator Increment has~:
\begin{itemize}
 \item no input data and one number output data, $d^o$  
\item a predicate execution function as the general case, for this case, because the operator does not have any input data, the function returns true for an ouput data does no contain a new token and false otherwise,
\item a process function that set the output data at the number of execution of this operator, 
\item an update function as the general case.
\end{itemize}
\subsubsection{$<$}
The operator $<$ has~:
\begin{itemize}
 \item two number input data, $d^i_0$ and $d^i_1$ and one boolean output data $d^o$, 
\item a predicate execution function as the general case ,
\item a process function that sets the output data at true for $a^t(d^i_0)<a^t(d^i_1)$ and false otherwise, 
\item an update function as the general case.
\end{itemize}
\subsubsection{Process}
The operator process has~:
\begin{itemize}
 \item one input data and one output data, 
\item a predicate execution function as the general case ,
\item a process function that can do everything, 
\item an update function as the general case.
\end{itemize}
\subsection{If/else composition pattern}
The if/else composition pattern has a petri net structure as the example $C_0$ and the figure~\ref{fig:evolution} shows its process. In this case, we execute the operator Process1 because the input data contained a boolean set at true. However, for a boolean set at false, the process2 would be executed. Therefore, this composition is the pattern for the condition if/else where an interpreted code seems like that~:
\begin{lstlisting}
if(a(d_0)==true)
  a(d_6)=Process1(a(d_1));
else
  a(d_6)=Process2(a(d_1));
\end{lstlisting} 
\begin{figure}
\begin{center}
\begin{tabular}{cc}
\includegraphics[height=3cm]{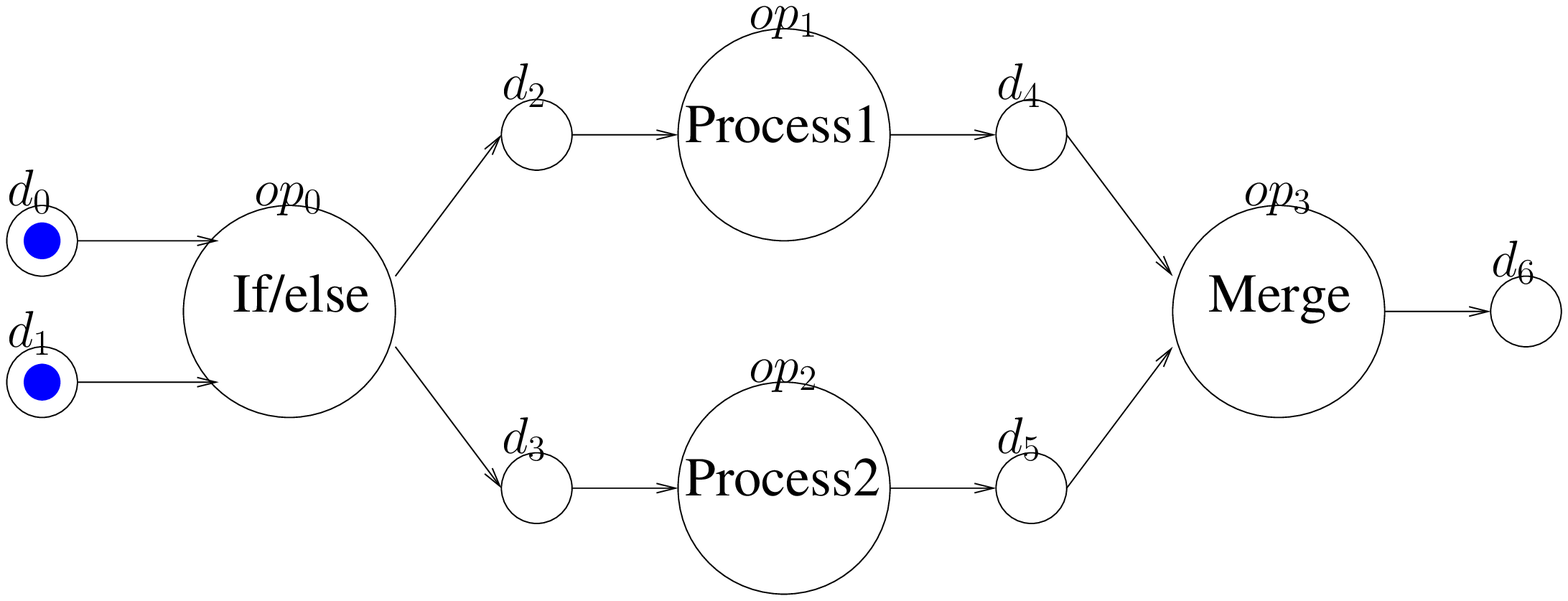}&\includegraphics[height=3cm]{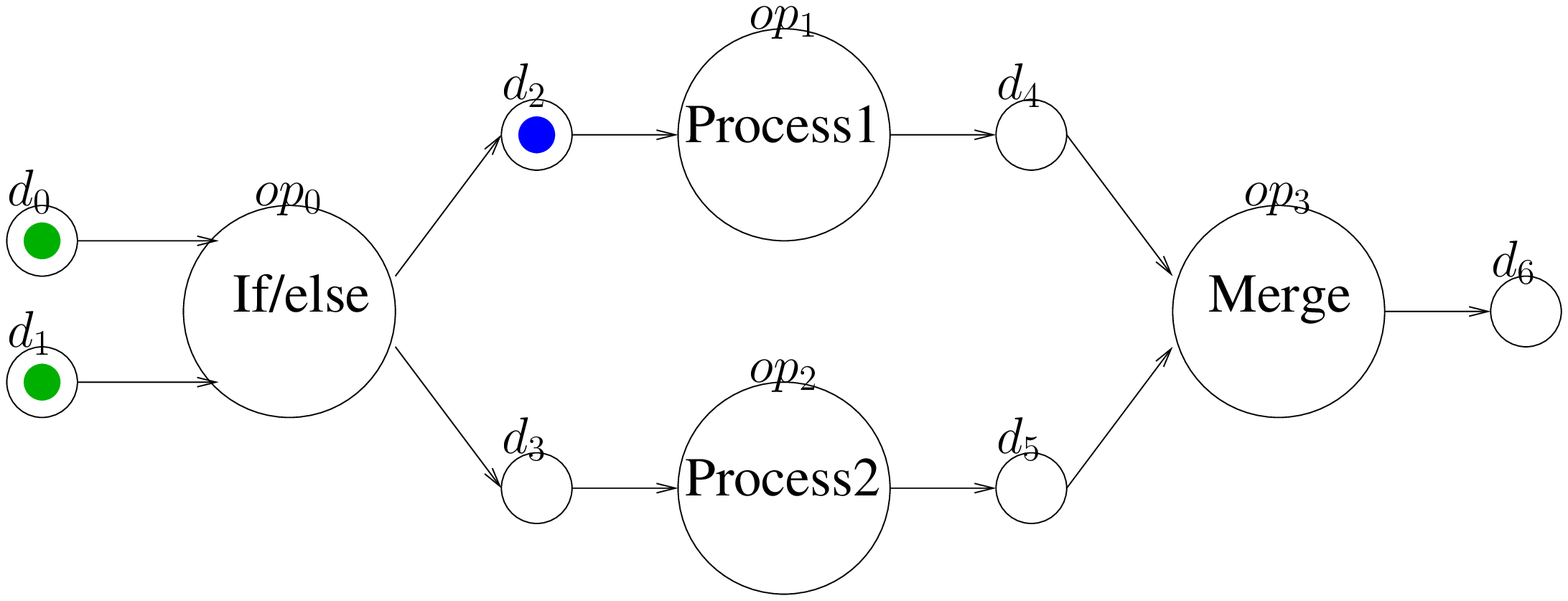}\\
step 0 & step 1\\\
\includegraphics[height=3cm]{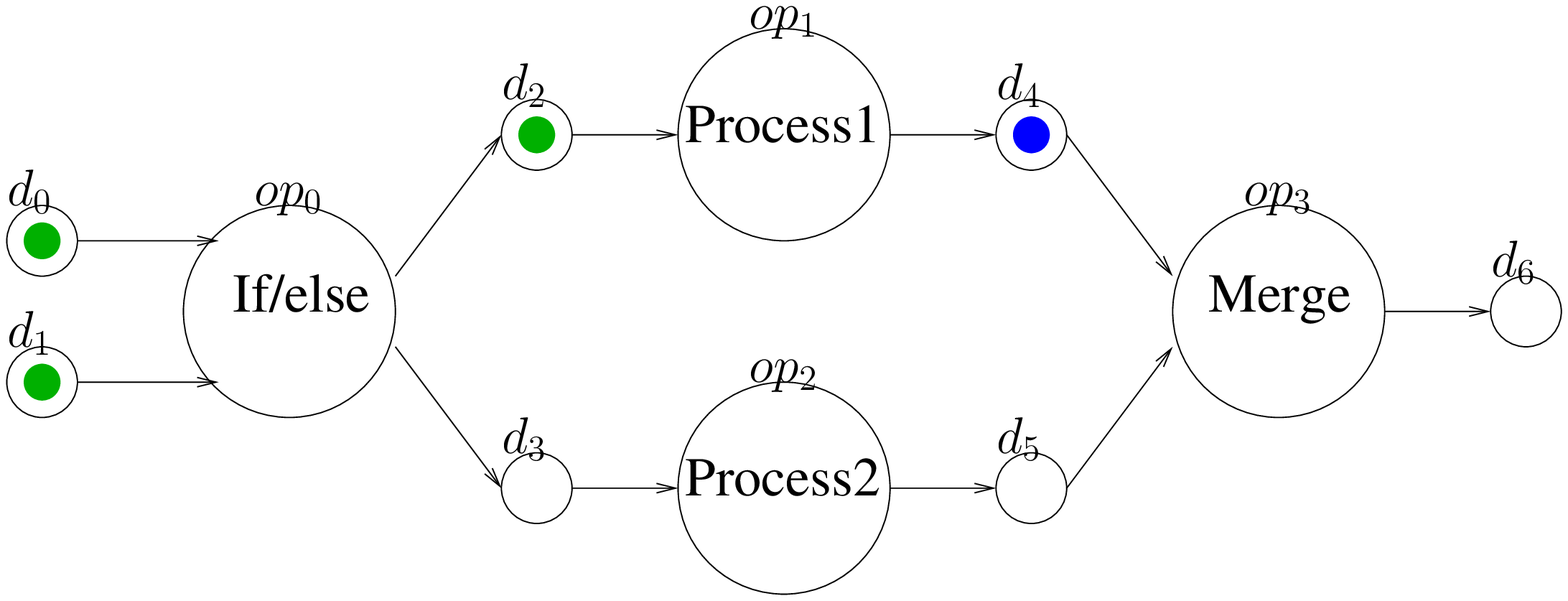}&
\includegraphics[height=3cm]{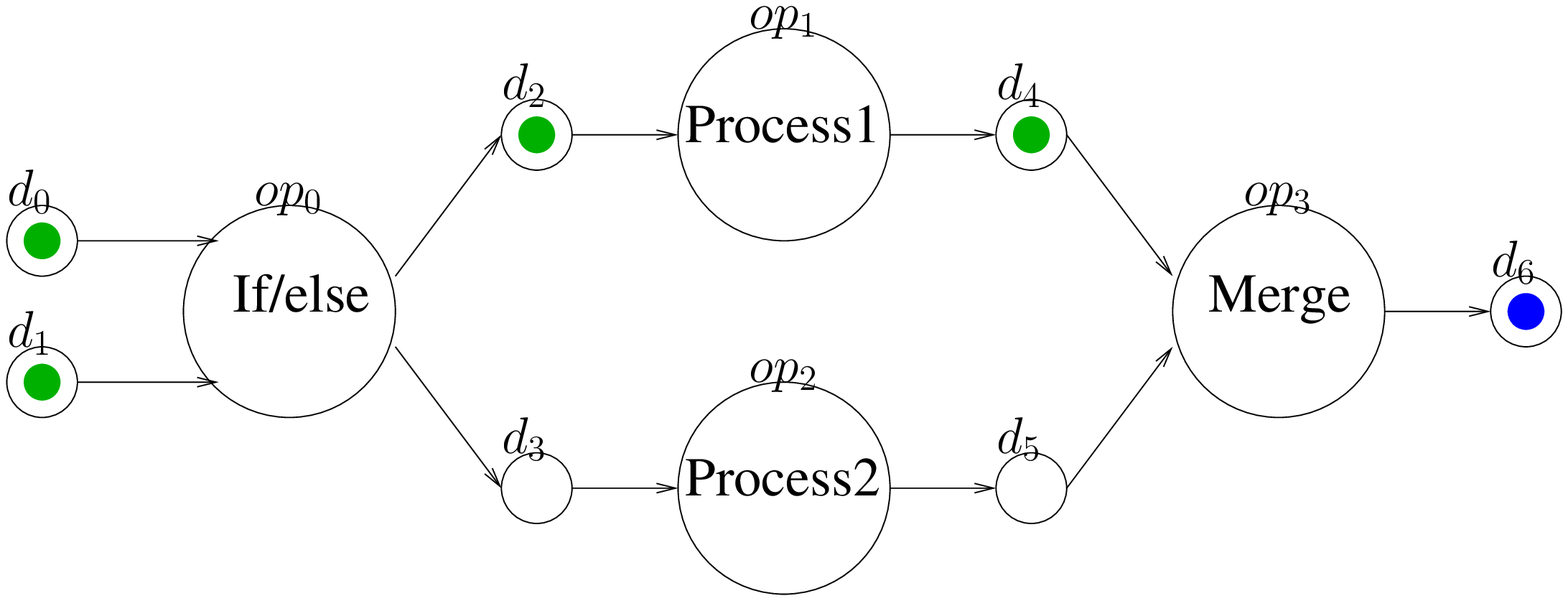}\\
step 2 & step 3
\end{tabular}
\caption{Execution process of the if/else composition pattern where the boolean data $d_0$ is set at true~: in step 1, we execute the operator if/else. Because the boolean data $d_0$ is set at true, we copy the input data $d_0$ to the output data in the if branch; in step 1, we execute the operator Process; in the step 2,  the operator Merge can be executed even if there is no token in one of the input data due to its special predicate execution function}
\label{fig:evolution}
\end{center}
\end{figure}
\subsection{Loop composition pattern}
\label{annex:loop}
The loop composition pattern has a petri net structure as the example $C_1$ and the figure~\ref{fig:loopprocess} shows its process. We can observe that we iterate the operator process. The key component in this composition is the operator synchrone allowing to wait the end of the loop before to start a new one. This composition is the pattern for the loop where an interpreted code seems like that~:
\begin{lstlisting}
for(int i=1;i<a(d_0);i++)
{
    a(d_3)=Process1(a(d_3));
}  
a(d_9)=a(d_3);
\end{lstlisting}
\begin{figure}
\begin{center}
\begin{tabular}{ccc}
\hspace{-2cm}\includegraphics[height=3cm]{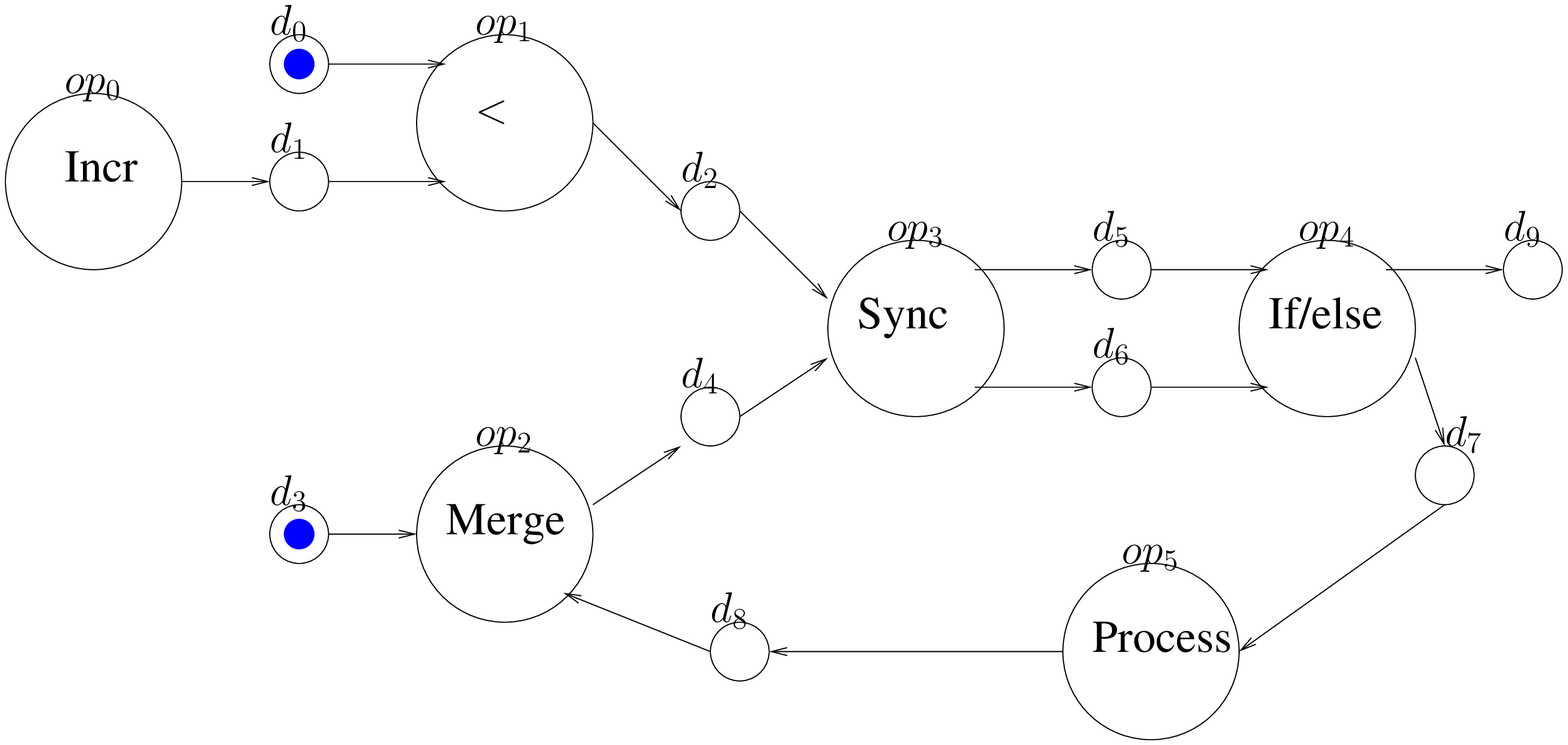}&\includegraphics[height=3cm]{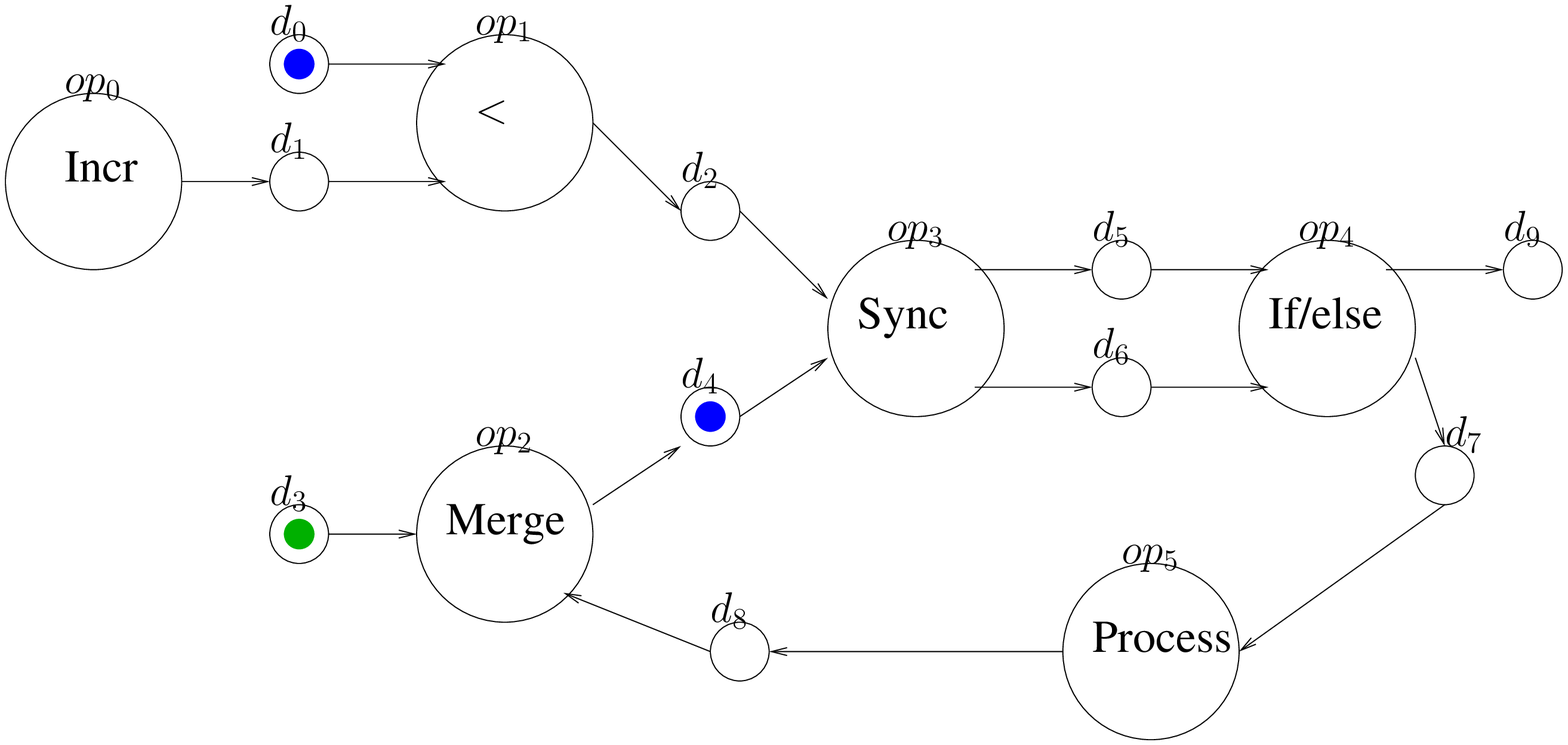}&\includegraphics[height=3cm]{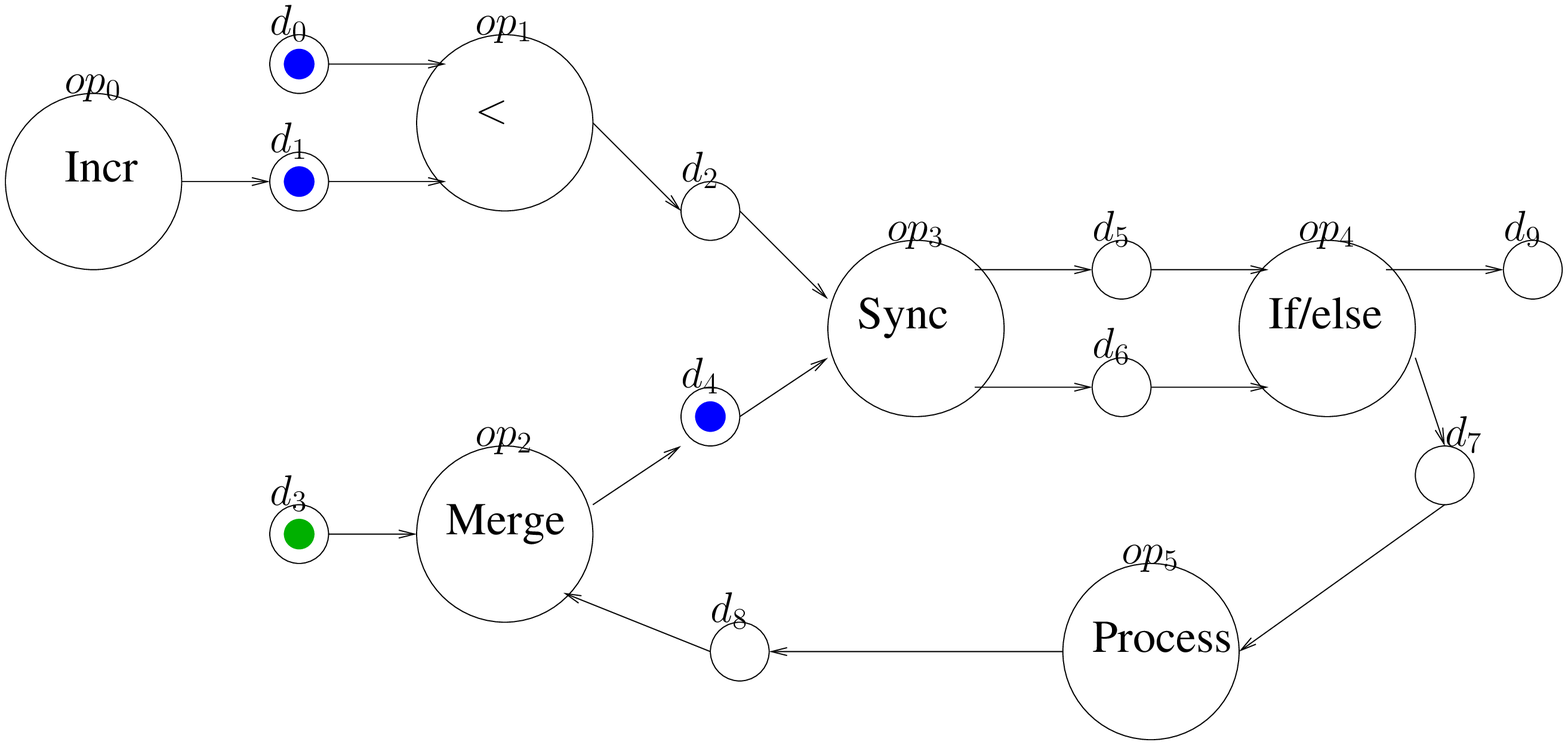}\\
step 0 & step 1 & step 2 \\
\hspace{-2cm}\includegraphics[height=3cm]{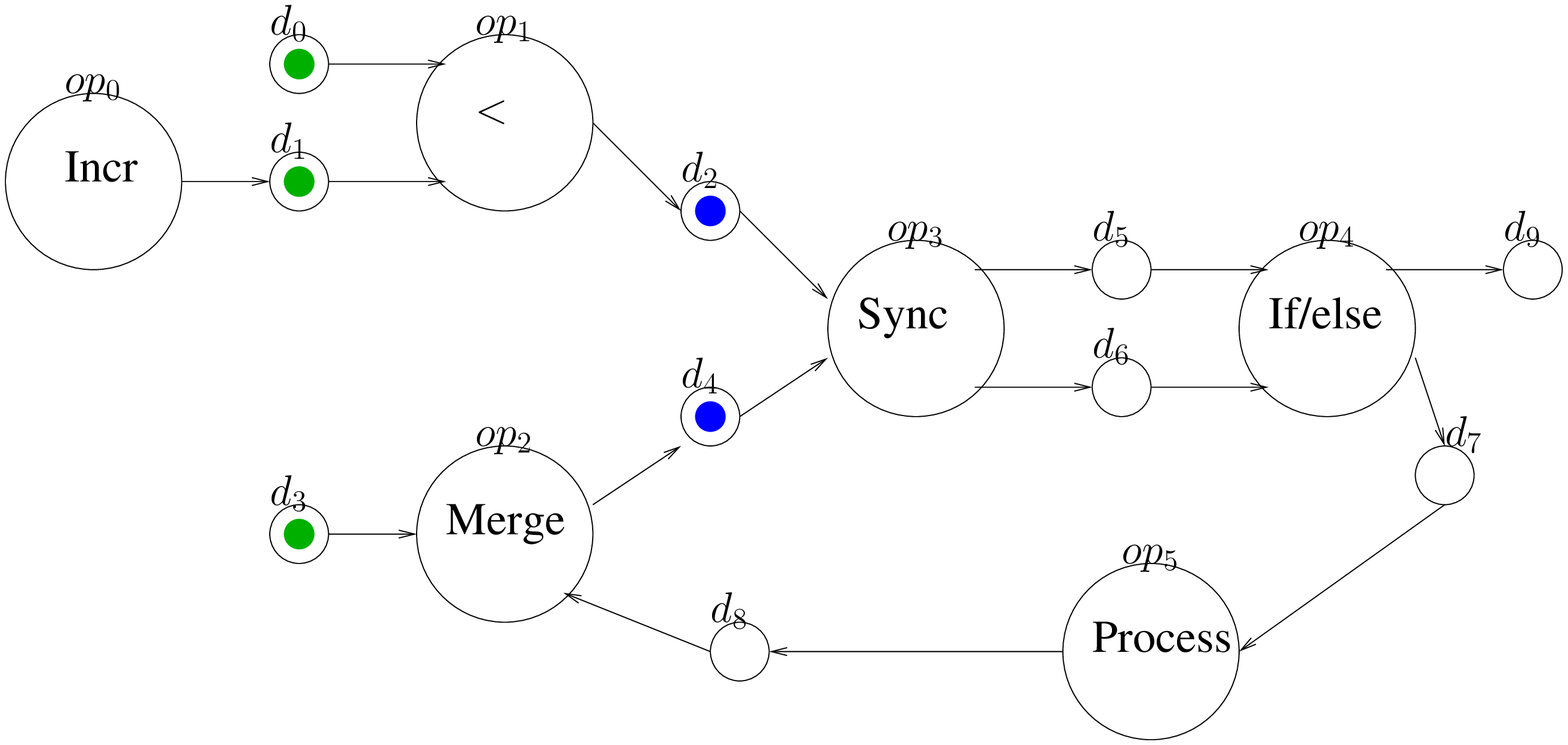}&\includegraphics[height=3cm]{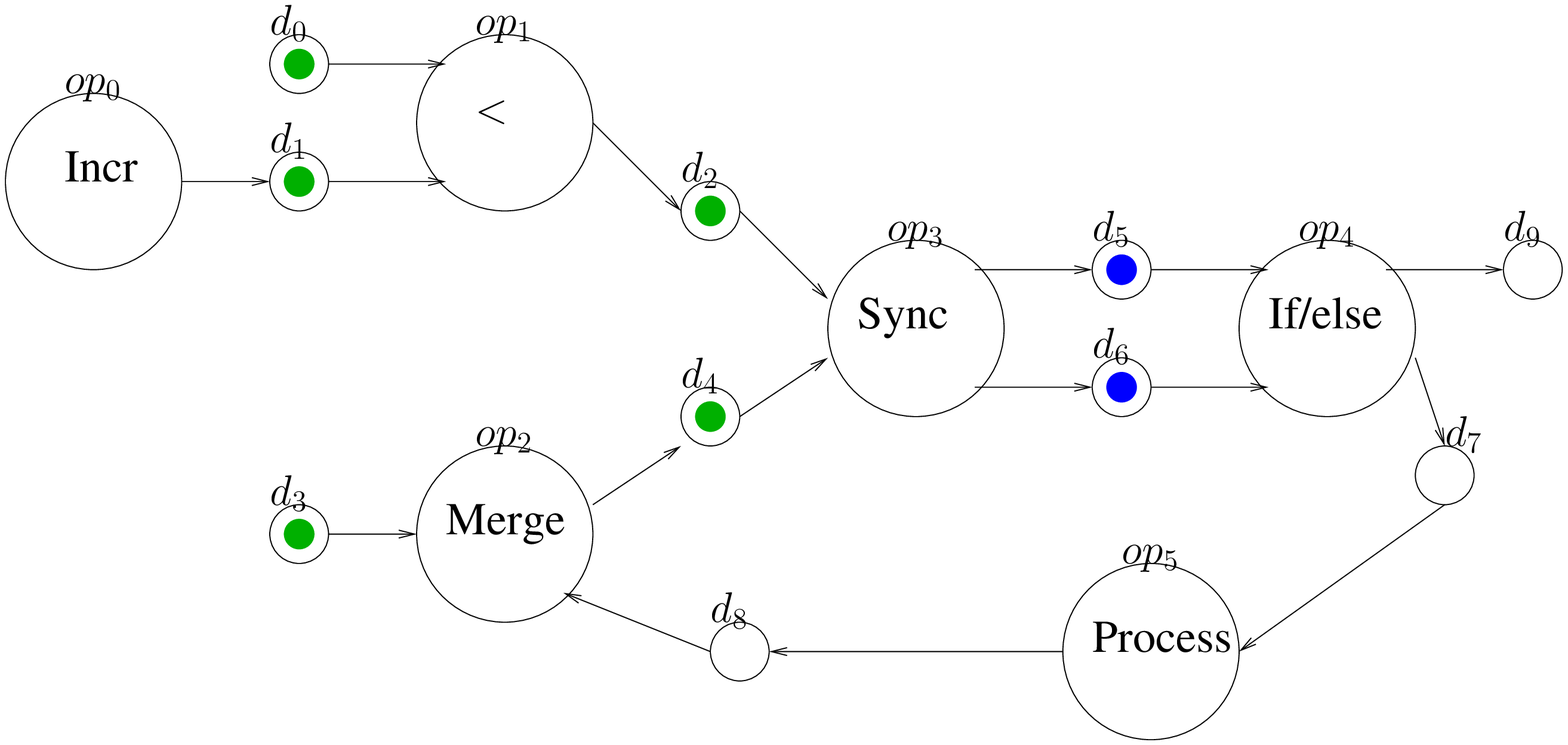}&\includegraphics[height=3cm]{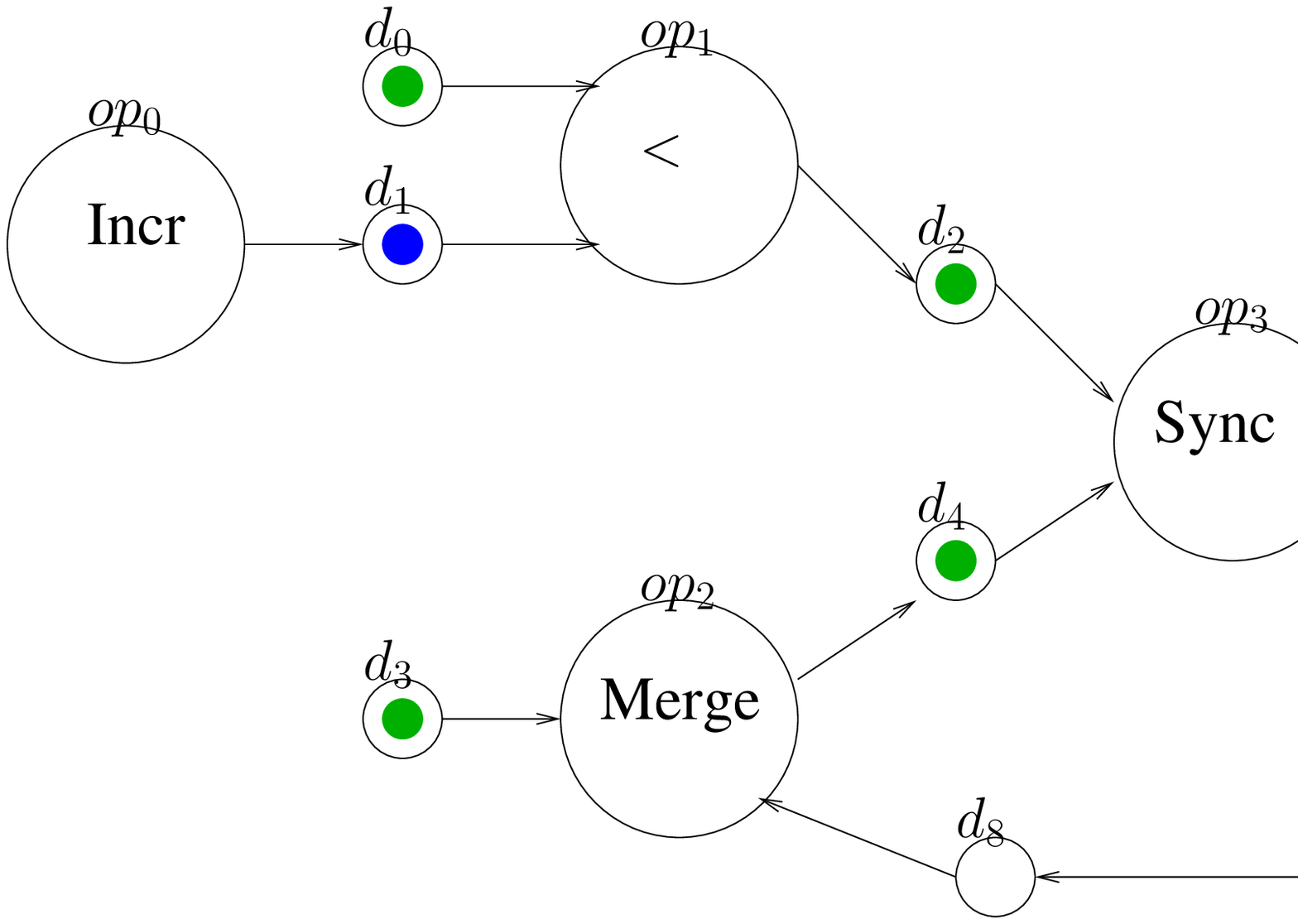}\\
step 3 & step 4 & step 5 \\
\hspace{-2cm}\includegraphics[height=3cm]{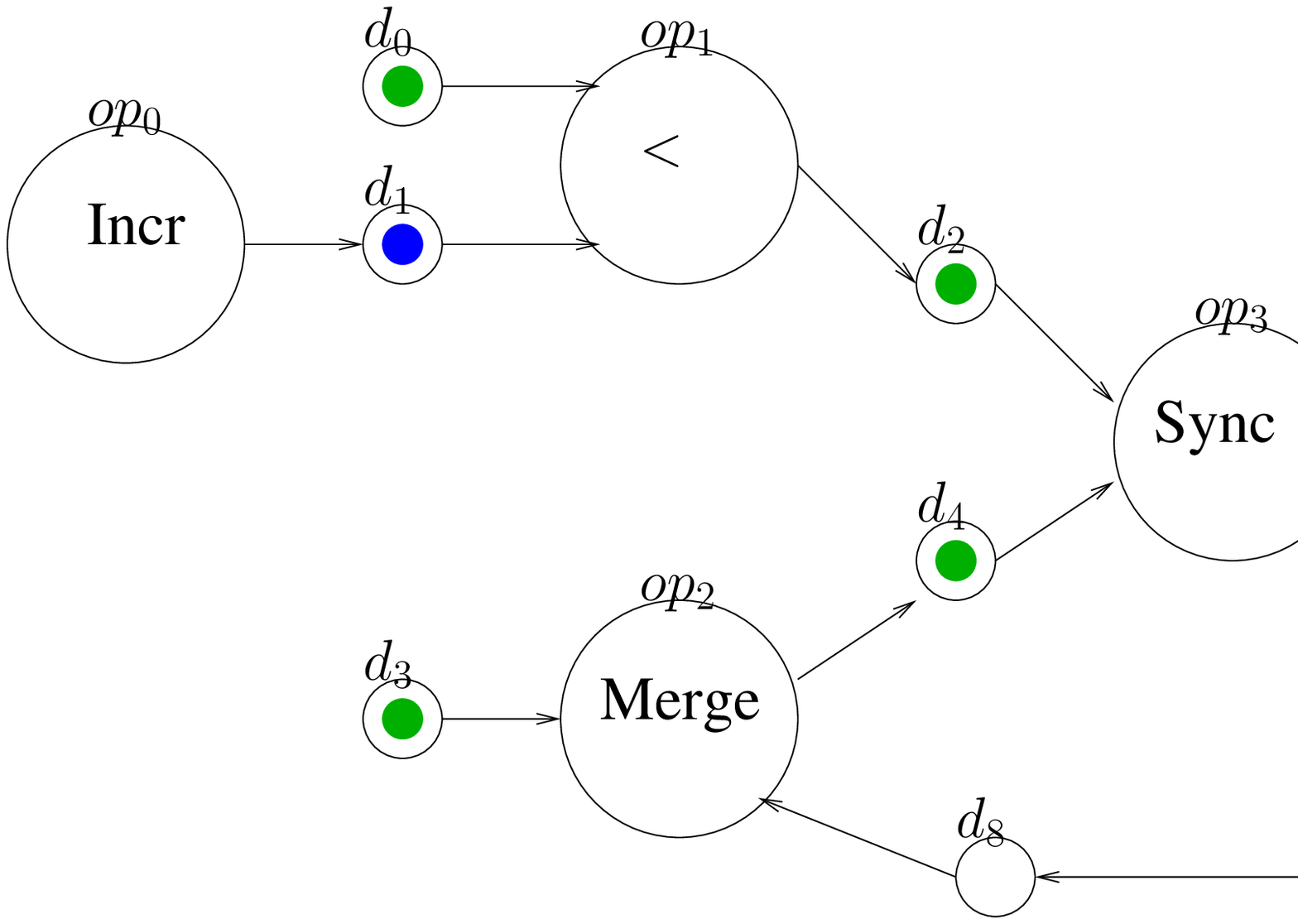}&\includegraphics[height=3cm]{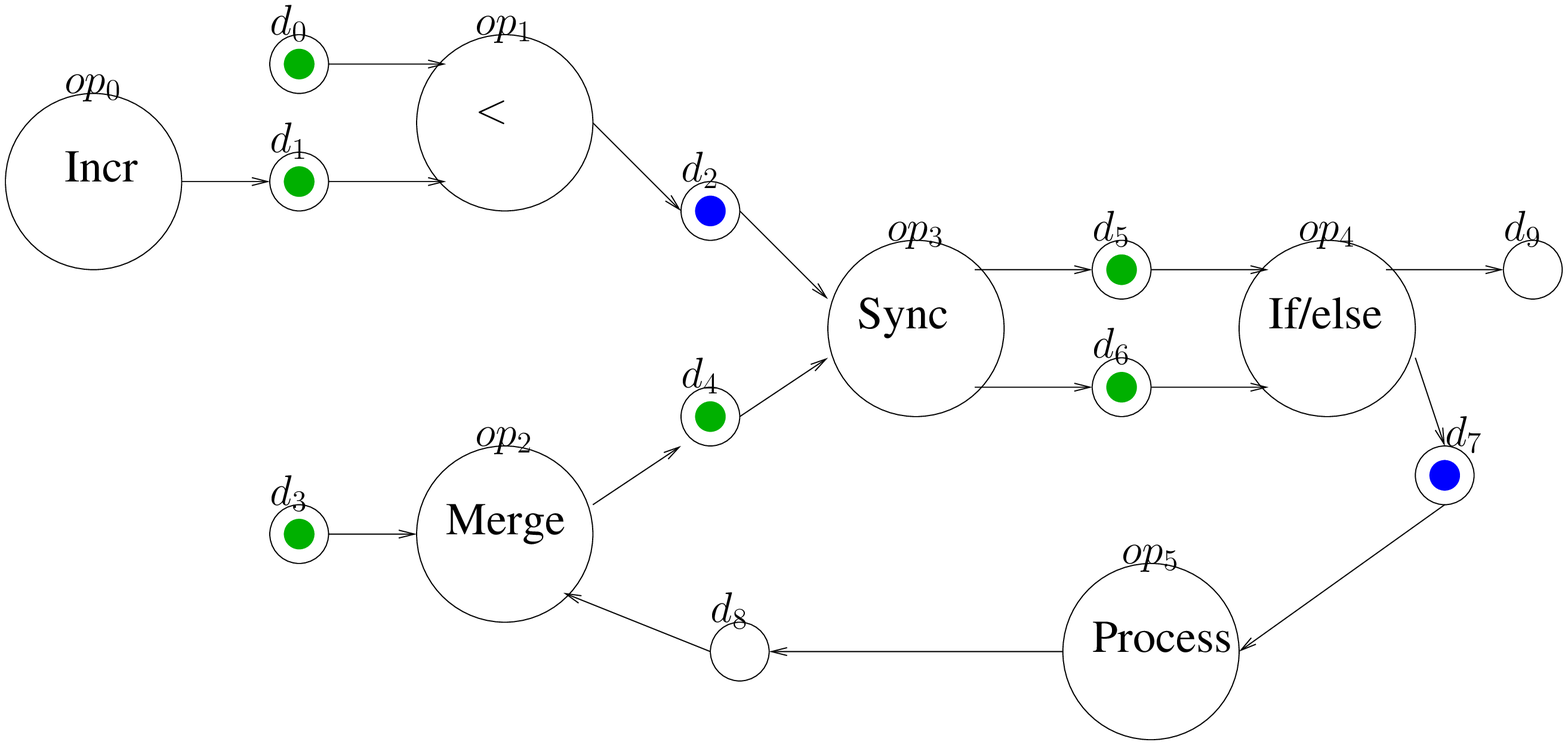}&\includegraphics[height=3cm]{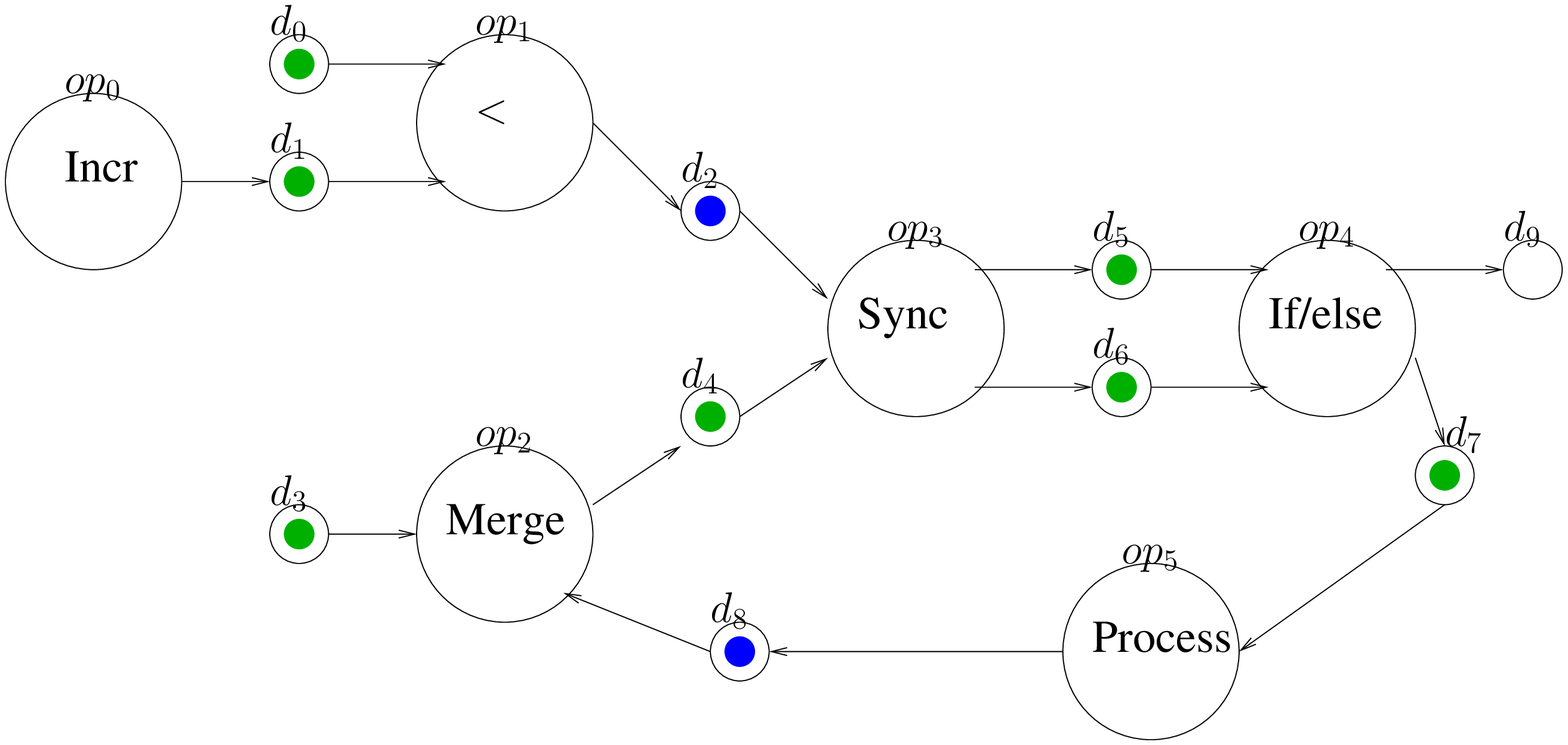}\\
step 6 & step 7 & step 8 \\
\hspace{-2cm}\includegraphics[height=3cm]{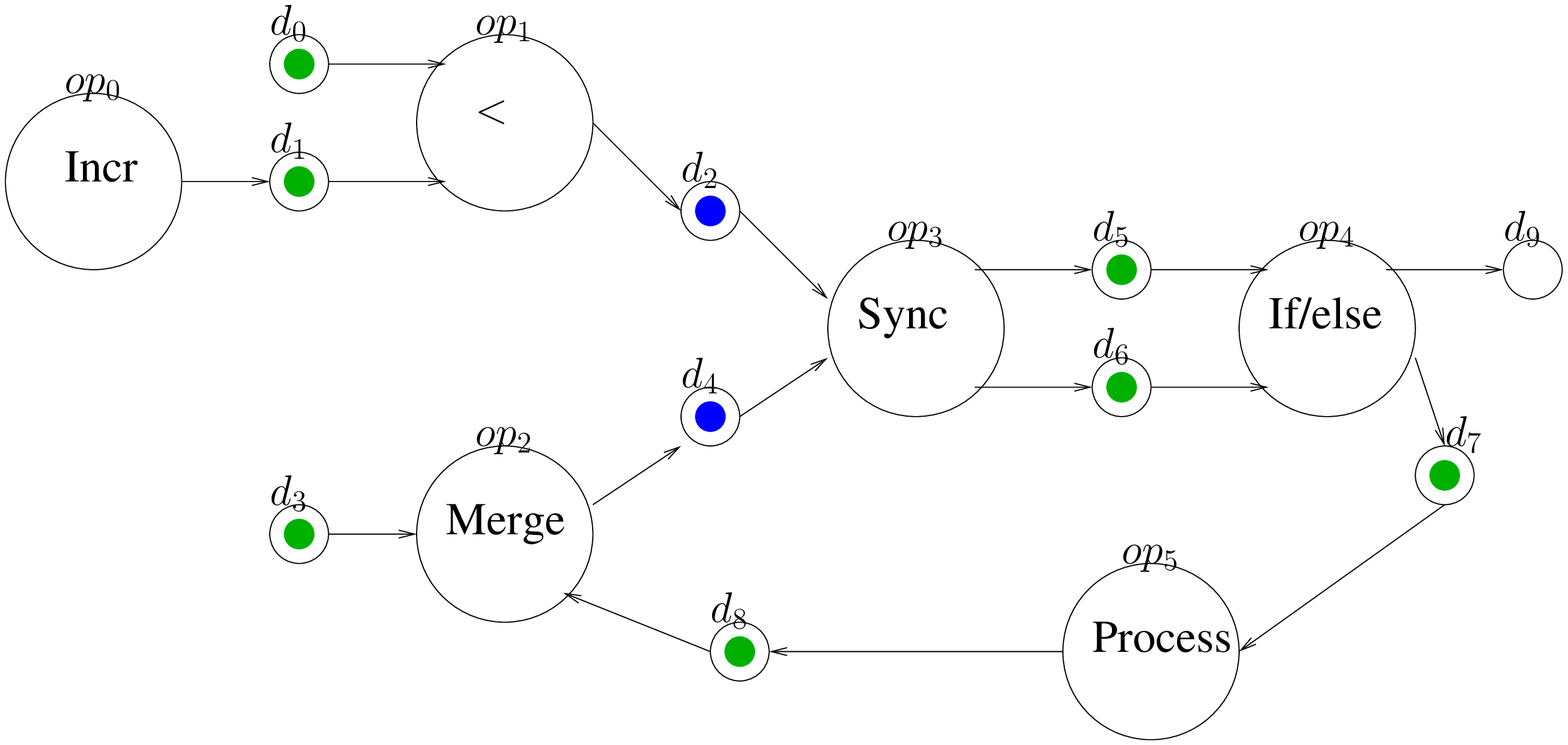}&\includegraphics[height=3cm]{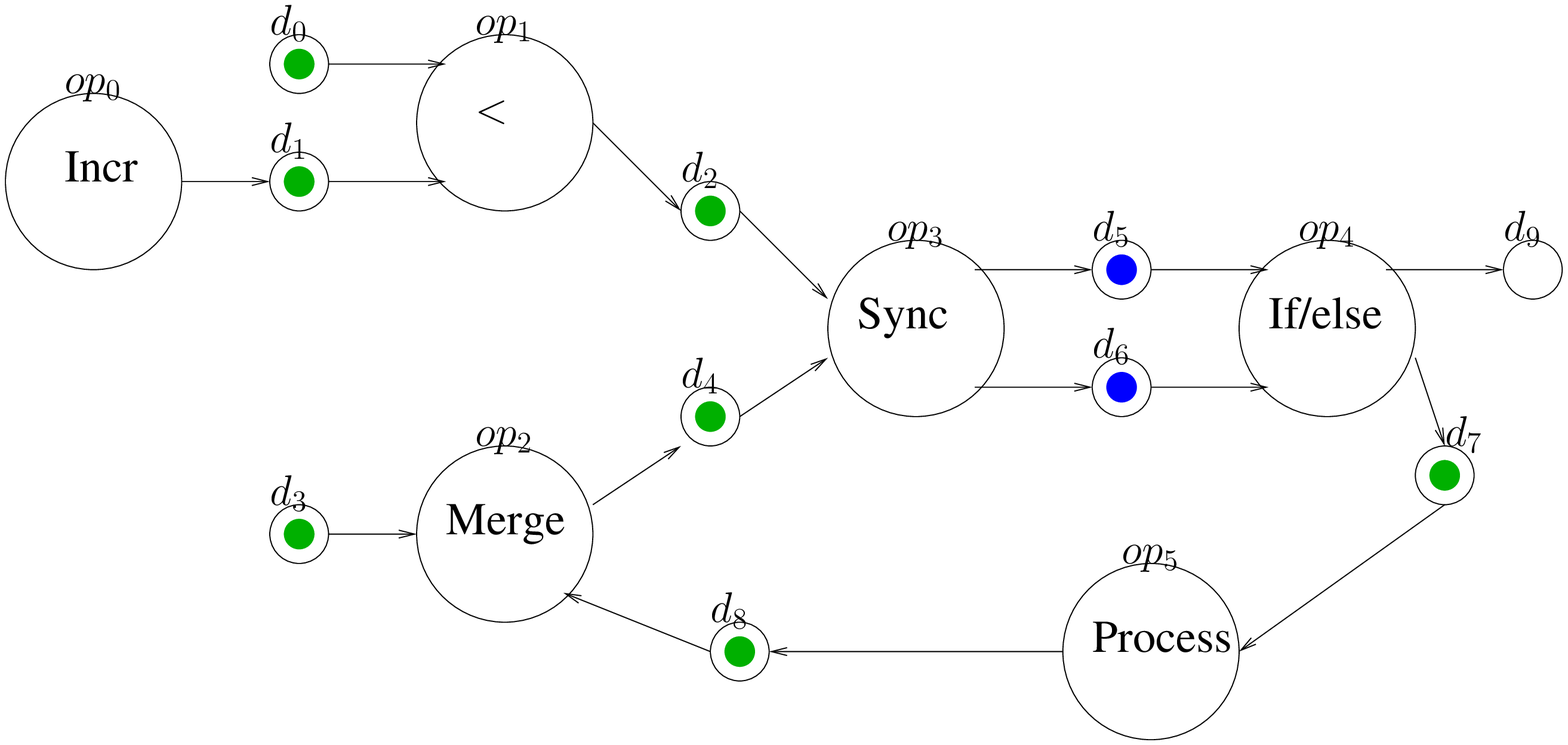}&\\
step 9 & step 10 &  
% \includegraphics[height=3cm]{petriloopname2.eps}&\includegraphics[height=3cm]{petriloopname3.eps}\\
% step 3 & step4 \\
% \includegraphics[height=3cm]{petriloopname4.eps}&\includegraphics[height=3cm]{petriloopname5.eps}\\
% step 3 & step4\\
% \includegraphics[height=3cm]{petriloopname6.eps}&\includegraphics[height=3cm]{petriloopname7.eps}\\
% step 3 & step4 \\
% \includegraphics[height=3cm]{petriloopname8.eps}&\includegraphics[height=3cm]{petriloopname9.eps}\\
% step 3 & step4 \\
% \includegraphics[height=3cm]{petriloopname10.eps}&\\
% step 3 & step4 \\
\end{tabular}
\caption{Execution process of the loop pattern composition where the number data $d_0$ is set at 10~: in step 0 two operators, Incr and Merge, can be executed and we make the choice to execute the operator Merge; in step 1 the Incr operator is executed that sets the output data at $1$ equal to its number of execution from the beginning; in step 2 the operator $<$ is executed  that sets the output data $a^t(d_2)$ at true since  $a^t(d_1)=1<a^(d_0)=10$; in step 3 two operators, Incr and Sync, can be executed and we choose the operator Sync; in step 4 two operators, Incr and if/else, can be executed but the operator Incr waits to be executed for longer time, its execution  sets the output data at $2$, $a^{t+1}(d_1)=2$, equal to its number of execution from the beginning; in step 5 two operators, if/else and $<$, can be executed but the operator if/else waits to be executed for longer time, its execution copy the input data , $a^t(d_6)$, to the output data $a^{t+1}(d_7)$, since the input boolean data $a^t(d_6)$ is set at true; in step 6 two operators,  $<$ and Process1, can be executed but the operator $<$ waits to be executed for longer time, its execution sets the output data $a^{t+1}(d_2)$ at true since  $a^t(d_1)=2<a^t(d_0)=10$; in step 7 the operator Sync cannot be executed due to its special predicate execution function, the operator Process is executed; in  step 8 the operator Merge is executed; in step 9 the operator Sync is executed; we iterate this loop 9 times until the operator $<$ set at false its output data}
\label{fig:loopprocess}
\end{center}
\end{figure}
\section{Conclusion}
In this article, we present how a composition is executed by the processor.  This execution depends globally on the connectivity between the operators and data and locally on the execution rules of each operator.  These execution rules are~: 1) a predicate executing function that returns true if the operator can be executed and false otherwise% where in the general case true for each input data containing an old or new token with at least one at new and each output data not containing a new token, 
, 2) a process that set the information of the output data depending on the information in the input data, 3) an update function that modifies the marking function for the input/output data  of the executed operator.%, in the general case, each input data  has an old token and each output data has an new token.  
 At each iteration, the processor executes the operator waiting to be executed for the longer time by the application of its process function and its update function. We iterate this process until all none of the operators cannot be executed. The processor component of the Cameleon language interprets a composition like this formalism. We demonstrate the possibilities of this formalism by the design of two fondamentals composition patterns,  loop and if/else statements, in a simple way.
% 
% We introduce three operators with special rules for the predicate executing function and the update functio
% n. Thansk to their,  We implement the processor of the cameleon language as this formalism with a modern architecture allowing an esay tune of the execution rules of your home-made operators.

%  the communication between an information processing system (such as a computer), and the outside world   
% However, a non-determinist action from outside can add/remove a single token in the data elements.
%   \\    

% For the description of this sequentiel process, we introduce a discrete time variable $t$ where, at each time increment, the marking is modified by~:
% \begin{enumerate}
%  \item the update after the operator execution,  $\mu^{t+1}(d) = u(d,\beta(op),I,0,\mu^t)$
% \item a token is added or removed  by an external user, $\mu^{t+1}(d) = m(\mu^t)$
% \end{enumerate}
% 
% \subsubsection{Evolution}
% The evolution is just the iteration of the unit execution.\\

%  In the application section, we will define some fundamental executing types to allow the simple implementation of a wide range of conditional exections and repetition. 

\end{document}